\documentclass[aps,prb,twocolumn,floatfix,superscriptaddress]{revtex4-1}

\usepackage{graphicx}
\usepackage[colorlinks,citecolor=magenta,linkcolor=blue]{hyperref}
\usepackage{txfonts}

\begin{document}

\title{ARPES/STM study of the surface terminations and 5$f$-electron character in URu$_2$Si$_2$}
\author{W. Zhang}


\author{H. Y. Lu}

\author{D. H. Xie}
\affiliation{Science and Technology on Surface Physics and Chemistry Laboratory, Mianyang 621908, China}

\author{W. Feng}
\author{S. Y. Tan}
\author{Y. Liu}
\author{X. G. Zhu}
\affiliation{Science and Technology on Surface Physics and Chemistry Laboratory, Mianyang 621908, China}
\author{Y. B. Huang}
\affiliation{Shanghai Institute of Applied Physics, CAS, Shanghai, 201204, China}

\author{Y. Zhang}
\author{Q. Q. Hao}
\affiliation{Science and Technology on Surface Physics and Chemistry Laboratory, Mianyang 621908, China}

\author{X. C. Lai}
\email[Corresponding author: ]{laixinchun@caep.cn}
\affiliation{Science and Technology on Surface Physics and Chemistry Laboratory, Mianyang 621908, China}

\author{Q. Y. Chen}
\email[Corresponding author: ]{chenqiuyun@caep.cn}
\affiliation{Science and Technology on Surface Physics and Chemistry Laboratory, Mianyang 621908, China}


\date{\today}

\begin{abstract}
Hidden order in URu$_2$Si$_2$ has remained a mystery now entering its 4th decade. The importance of resolving the nature of the hidden order has stimulated extensive research.
 Here we present a detailed characterization of different surface terminations in URu$_2$Si$_2$ by angle-resolved photoemission spectroscopy, in conjunction with scanning tunneling spectroscopy and DMFT calculations that may unveil a new piece of this puzzle. The U-terminated surface is characterized by an electron-like band around the $\bar{X}$ point, while a hole-like band for the Si-terminated surface. We also investigate temperature evolution of the electronic structure around the $\bar{X}$ point from 11 K up to 70 K, and did not observe any abrupt change of the electronic structure around the coherence temperature (55 K). The $f$ spectral weight gradually weakens upon increasing temperature, still some $f$ spectral weight can be found above this temperature. Our results suggest that surface terminations in URu$_2$Si$_2$ are an important issue  to be taken into account in future work.

\end{abstract}

\maketitle

\begin{figure*}
\includegraphics[width=15cm]{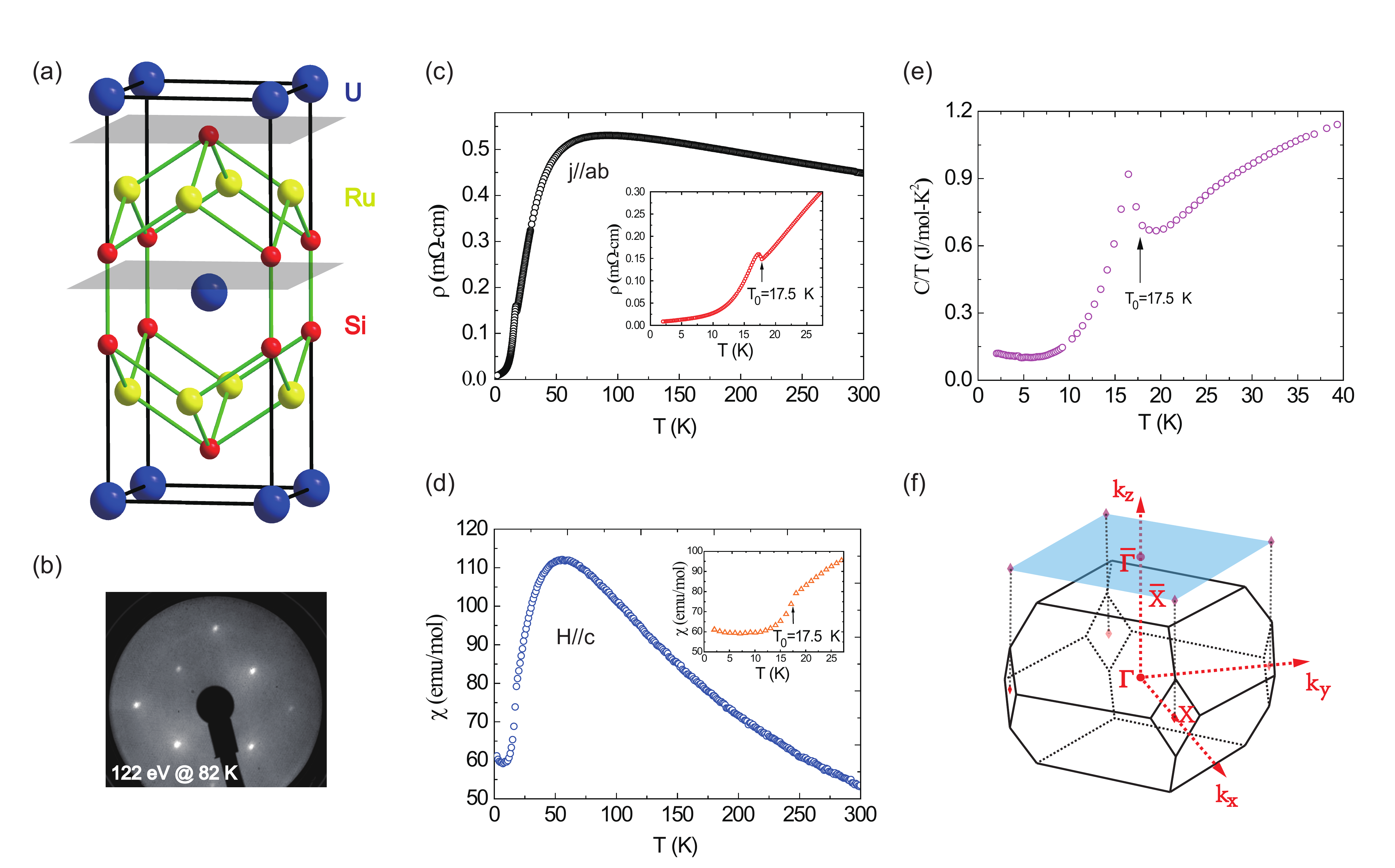}
\caption{ (a) Crystal structure of URu$_2$Si$_2$. Possible cleaving planes have been marked gray.  (b) LEED patterns of the cleaved surface measured at 82 K. (c) Temperature dependence of the electrical resistivity. (d) Dc-magnetic susceptibility $\chi$(T) measured in the magnetic field of 0.1 T parallel to the $c$-axis. (e) Temperature dependence of the specific heat. (f) Brillouin zone of the bulk URu$_2$Si$_2$, and the projected (001) surface Brillouin zone has been marked blue. }
\end{figure*}

\begin{figure}
\includegraphics[width=8cm]{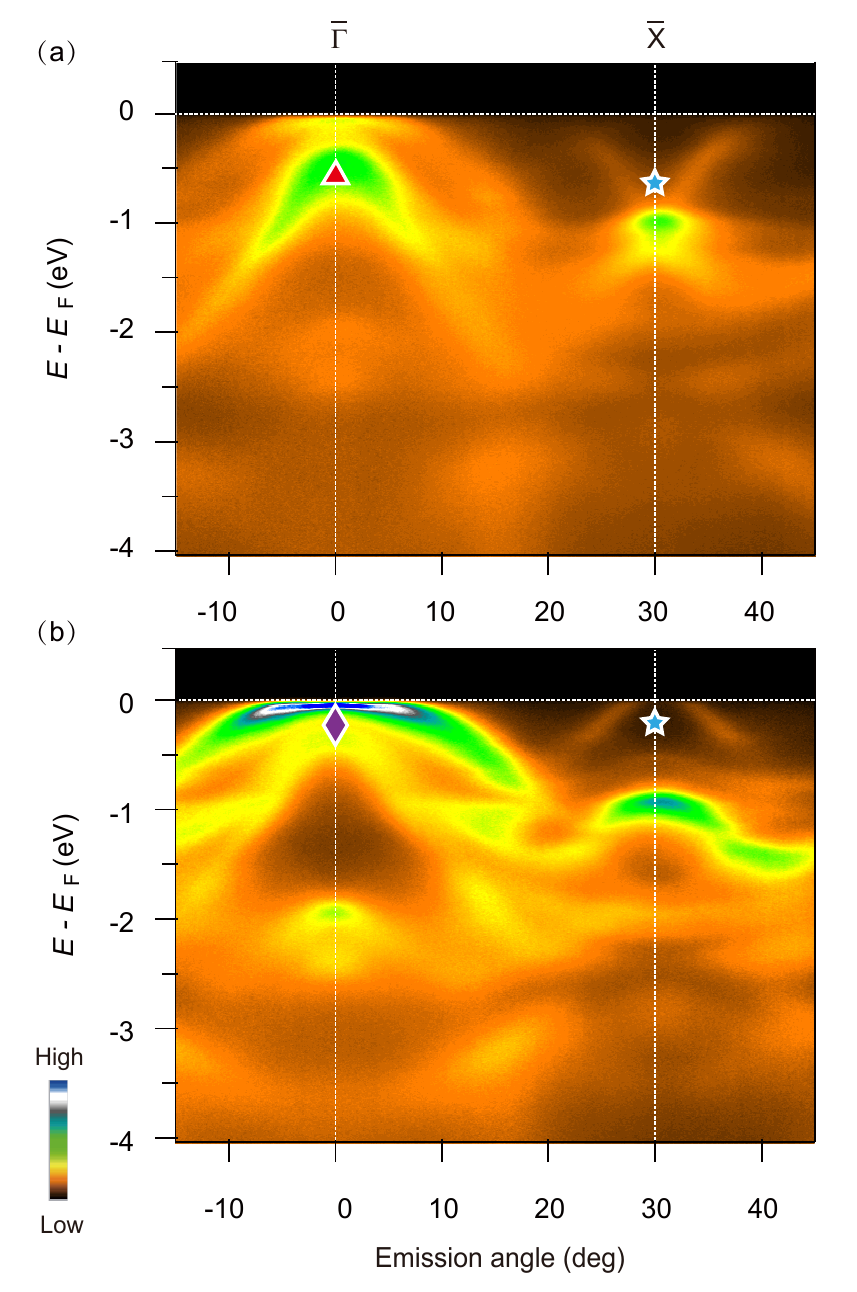}
\caption{ARPES characterization of different surface terminations in URu$_2$Si$_2$ at 82 K.  (a) ARPES data of the U-terminated surface. (b) ARPES data of the Si-terminated surface. The data were taken along the $\bar{\Gamma}$-$\bar{X}$ direction at 82 K. For the meaning of the symbols see the main text.}
\end{figure}

\section{Introduction}

``Hidden order" in URu$_2$Si$_2$, one of the most mysterious challenges in current condensed matter physics, has attracted extensive research during the past thirty years \cite{Palstra.85,Maple.86,Schmidt.10,Aynajian10383,Andres.09,Mydosh.11}. This second-order phase transition at $T_0$=17.5 K is marked by a jump in specific heat and removal of a significant fraction of total entropy \cite{Palstra.85,Maple.86}. While the small antiferromagnetic moment of 0.03 $\mu_B$ detected by neutron scattering experiment was too small to explain this large entropy release upon the transition \cite{PhysRevLett.58.1467}. Many theoretical models have been proposed to explain this mysterious order \cite{PhysRevLett.103.107202,PhysRevB.83.165110,PhysRevB.82.205103,PhysRevLett.106.086401,PhysRevLett.106.106601,PhysRevLett.106.196407,Chandra.13}. Among them, a key question is the property of Fermi surface gapping and instability in the momentum space \cite{Maple.86,PhysRevLett.98.016401,PhysRevLett.99.116402,Wiebe.07,Andres.14}.


Angle-resolved photoemission spectroscopy (ARPES) is a powerful tool to directly observe the Fermi surface topology and electronic structure in the momentum space in solid materials. Earlier pioneering ARPES studies of URu$_2$Si$_2$ have concentrated on the paramagnetic phase  and established the existence of hole pockets at the $\bar{\Gamma}$, $\bar{Z}$ and $\bar{X}$ points of the Brilloune zone \cite{Ito.99,Denlinger.01,Denlinger.02}. Subsequent ARPES results proposed either a heavy band collapsing towards the Fermi level ($E_F$) through the transition \cite{Andres.09} or a heavy band developing below $E_F$ \cite{Yoshida.12}. In contrast, the existence of weakly dispersive states was observed not to shift from above to below $E_F$, and these states rapidly hybridize with conduction bands upon entering the hidden order phase \cite{PhysRevLett.110.186401}. Three dimensional nature of the Fermi surface in URu$_2$Si$_2$ has also been obtained \cite{Kawasaki.11,Meng.13}. These ARPES results have shed new light on the ``hidden order" problem and revealed important aspects of this mystery, but there is still much work left to be done.
\begin{figure*}
\includegraphics[width=18cm]{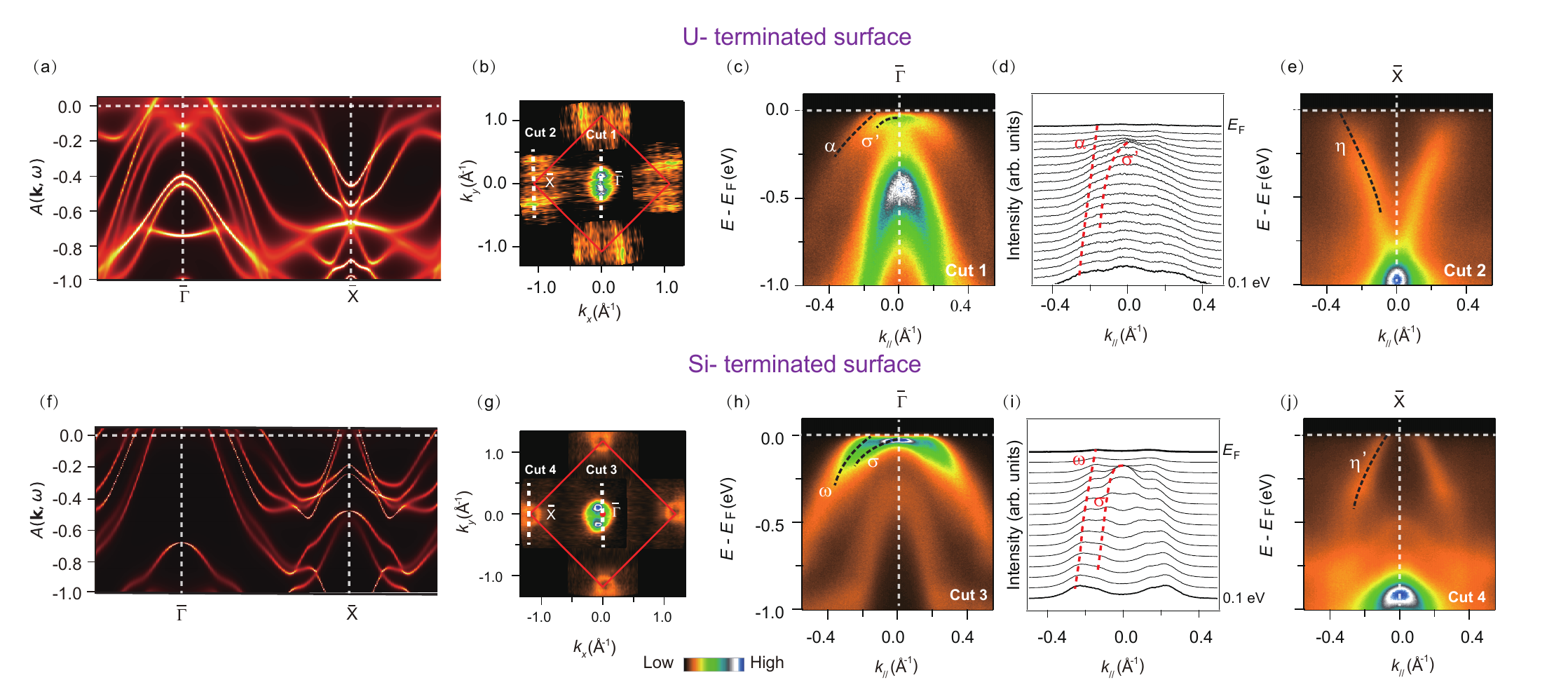}
\caption{ Band structure of the U- and Si- terminated surfaces measured at 82 K. (a) Calculated band structure of the U- terminated surface. (b) Photoemission intensity map of the U- terminated surface at $E_F$ integrated over a window of ($E_F$-10 meV, $E_F$+10 meV). (c) Low energy band structure around the $\bar{\Gamma}$ point along cut1. (d) Momentum distribution curves (MDCs) around the $\bar{X}$ point along cut1 near the Fermi level. (e) Low energy band structure around the $\bar{\Gamma}$ point along cut2. (f-j) Same as (a-e), but for the Si- terminated surface. }
\end{figure*}

One of the important points to be understood is the surface termination in URu$_2$Si$_2$. Previous ARPES studies on the structurally similar RET$_2$Si$_2$ (RE=Ce, Eu, Yb and T=Co, Rh, Ir) materials have
indicated that these materials can be easily cleaved since the bond energy between RE atoms and the neighbouring Si-layer is much weaker than within the Si-T-Si layers, leading to either Si or RE- terminated surfaces, and two sets of band structure have been revealed for these compounds \cite{Patil.16,PhysRevB.75.045109,PhysRevLett.107.267601,PhysRevLett.105.237601,Hoppner.14}. URu$_2$Si$_2$ has the same crystal structure as the  RET$_2$Si$_2$ compounds, and it should also have different cleaved surfaces. Indeed, different cleaved surfaces have been observed by scanning tunneling spectroscopy (STM) in URu$_2$Si$_2$ \cite{Schmidt.10,Aynajian10383}. However, to the best of our knowledge, the electronic structure on different terminated surfaces has never been reported in URu$_2$Si$_2$ by ARPES \cite{Andres.14,Andres.13,Andres.09,Kawasaki.11,Yoshida.12,Yoshida.10,Dakovski.11,Ito.99,Meng.13,PhysRevLett.110.186401}, which is very important given the surface sensitivity of this technique.

Another important point is to understand the interplay of localization and itinerancy of the $f$ electrons. According to the standard model of heavy-fermion behavior, $f$ electrons are localized at high temperature, while their exchange coupling to conduction electrons leads to the formation of bands with heavy masses as temperature is lowered and the $f$ electrons become itinerant \cite{RevModPhys.48.219}. The detailed evolution of the localized-to-itinerant transition has been observed in some of the Ce- based systems \cite{Qiuyun.17,CeIrIn5_qiuyun,CeRhIn5_qiuyun}. Temperature-induced changes have also been observed in UPd$_2$Al$_3$, which is closely related to the localized-to-itinerant transition \cite{Fujimori.07}. By contrast, previous ARPES results  on URu$_2$Si$_2$  have concentrated on the relationship between the electronic structure and the ``hidden order" at low temperatures \cite{Andres.14,Andres.13,Andres.09,Kawasaki.11,Yoshida.12,Yoshida.10,Dakovski.11,Ito.99,Meng.13,PhysRevLett.110.186401}. One of the key results of these studies is that weakly dispersive states rapidly hybridize with light conduction bands just upon entering the hidden order phase around the $\bar{\Gamma}$ point  \cite{Andres.13}, which is in marked contrast to the gradual crossover behavior expected in Kondo lattice systems. Meanwhile, there is another key temperature scale in URu$_2$Si$_2$. The resistivity undergoes a rapid decrease below 55 K \cite{Palstra.85,Maple.86}, which is normally regarded as the beginning of the development of coherence between $f$ and conduction electrons.  However, the evolution of the electronic structure at this cross-over is still unknown and is an important element of the current study.

In this paper, we also present a detailed characterization of the different surface terminations in URu$_2$Si$_2$ by ARPES, STM/STS and DMFT calculations. The different terminations can be easily distinguished by an electron-like band around the $\bar{X}$ point for the U-terminated surface, while a hole-like band for the Si-terminated surface. Furthermore, temperature-dependent measurements were performed around the $\bar{X}$ point for the Si-terminated surface from 11 K up to 70 K, and we did not observe any abrupt change of the electronic structure around the coherence temperature. The $f$ spectral weight gradually weakens upon increasing temperature, still small part of $f$ spectral weight can be found above this temperature.

\section{Experimental and computational details}

Single crystals of URu$_2$Si$_2$ were grown by the Czochralski method in a tetra-arc furnace with a continuously purified Ar atmosphere and subsequently annealed at 900 $^\circ$C  under ultrahigh vacuum for 10 days. The electrical resistivity, magnetic susceptibility and specific heat measurements were performed using a Physical Property Measurement System (PPMS-9).
Samples were cleaved in-situ along the $c$ axis at a base pressure of better than $6\times10^{-11}$~mbar at 82 K. ARPES measurements were performed with SPECS UVLS discharge lamp (21.2 eV He-I¦Á light). All data were collected with Scienta R4000 electron analyzers. The overall energy resolution was about 15~meV or better, and the typical angular resolution was $0.2^{\circ}$. A freshly evaporated gold sample was used  to determine $E_F$. Temperature-dependent ARPES measurements were performed from high to low temperature. STM experiments were performed in an ultrahigh vacuum, low temperature STM apparatus with a base pressure of $2\times10^{-11}$~mbar. All topographic images were recorded in the constant current mode. STM chamber is connected with the ARPES chamber using a radical distribution chamber with a base pressure of $4\times10^{-11}$~mbar, so the samples can be transferred directly from ARPES chamber to STM chamber under ultra high vacuum conditions.

Our calculation includes two parts: a slab model calculation together with a fully self-consistent DFT+DMFT method to explore the electronic structure. For the slab calculation, 9 and 7 layers slab structures have been studied for the U- and Si- terminated surfaces, respectively. The vacuum thickness is chosen to be 10 {\AA} according to the convergence of total energy and band structure.
The electronic structure is carried out with the constructed U/Si based terminated surfaces without structure relaxation following ref. [42]. we tried to calculate the electronic structures of URu$_2$Si$_2$ with the combination of density functional theory and single-site dynamical mean-field theory (dubbed as DFT + DMFT)~\cite{Kotliar.06}. The DFT + DMFT method is probably the most powerful established approach to study the electronic structures of strongly correlated materials. It has been widely used to study the correlated $4f$ or $5f$ electron systems~\cite{PhysRevB.94.075132,Shim1615,shim:2007}. In the DFT part, the WIEN2K code was employed, which implements a full-potential linear augmented plane-wave formalism~\cite{wien2k}. The DFT calculations were done on a $13 \times 13 \times 2$ Monkhorst-Pack $k$-mesh, and the spin-orbit coupling is taken into account during calculation. We used $R_{\text{MT}}K_{\text{MAX}} = 7.0$ and $G_{\text{MAX}} = 9.0$, and chose the generalized gradient approximation (Perdew-Burke-Ernzerhof functional)~\cite{PhysRevLett.100.136406} to express the exchange-correlation potential. In the DMFT part, all the uranium atoms are treated to be equivalent. As for the Coulomb interaction only considering the correlation among the U-5$f$ orbitals, a four-fermion interaction matrix is built which is parameterized by the Slater integrals $F_k$.  The general interaction matrix was parameterized using the Coulomb interaction $U$ and the Hund's exchange $J$ via the Slater integrals. They were 6.0 eV and 0.6 eV, respectively.
The constructed multi-orbital Anderson impurity models were solved using the hybridization expansion continuous-time quantum Monte Carlo impurity solver (dubbed as CT-HYB). The calculated temperature is 82 K.
The convergence criteria for charge and energy were $10^{-4}$e and $10^{-4}$Ry, respectively.
The final output were Matsubara self-energy function $\Sigma$($i\omega_n$) and impurity Green's function $G$($i\omega_n$), which were then utilized to obtain the integral spectral functions $A(\omega)$ and momentum-resolved spectral functions $A$(\textbf{k},$\omega$).



\section{Results and discussions}

URu$_2$Si$_2$ crystallizes in the body-centered tetragonal ThCr$_2$Si$_2$-type structure, which belongs to the D4$h$ point group (space group $I4/mmm$ No. 139), as shown in Fig. 1(a). Temperature dependence of the electrical resistivity, dc-magnetic susceptibility for the $ab$-plane and the specific heat of our samples are displayed in Figs. 1(c, d) and 1(e), respectively, from which obvious hidden order transition can be observed at around 17.5 K. Brillouin zone of the bulk URu$_2$Si$_2$, and the projected (001) surface Brillouin zone have been displayed in Fig. 1(f). After cleavage, sharp low energy electron diffraction (LEED) patterns can be observed in Fig. 1(b) without any surface reconstruction.
\begin{figure*}
\includegraphics[width=17cm]{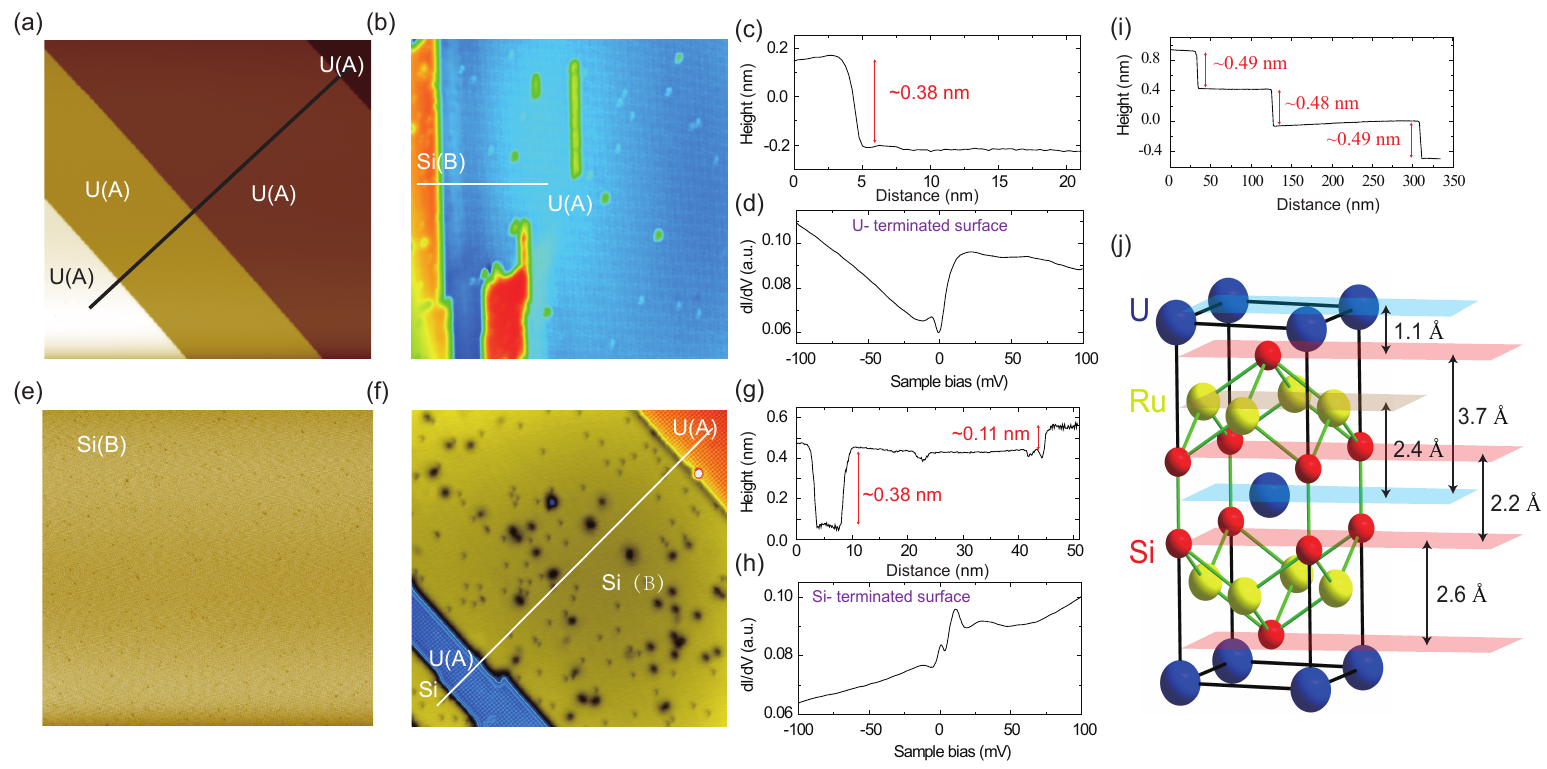}
\caption{STM topography and spectroscopy on URu$_2$Si$_2$ at 4.5 K. (a, b) Constant current topographic image (20 mV, 60 pA) over a 300$\times$300 nm$^2$ (a), 50$\times$50 nm$^2$ (b) area of the U-dominated surfaces. (c) The relative heights between U- and Si-terminated surfaces in (b). (d) Averaged electronic density of states (DOS) of the U-terminated surface. (e, f) Constant current topographic image (20 mV, 60 pA) over a 300$\times$300 nm$^2$ (e), 40$\times$40 nm$^2$ (f) area of the Si-dominated surfaces. (g) The relative heights between U- and Si- surfaces in (f). (h) Averaged electronic density of states (DOS) of the Si-terminated surface. (i) The relative heights between U-terminated surfaces in (a)}. (j) Schematic diagram illustrating the height of different layers in URu$_2$Si$_2$.
\end{figure*}

\begin{figure*}
\includegraphics[width=18cm]{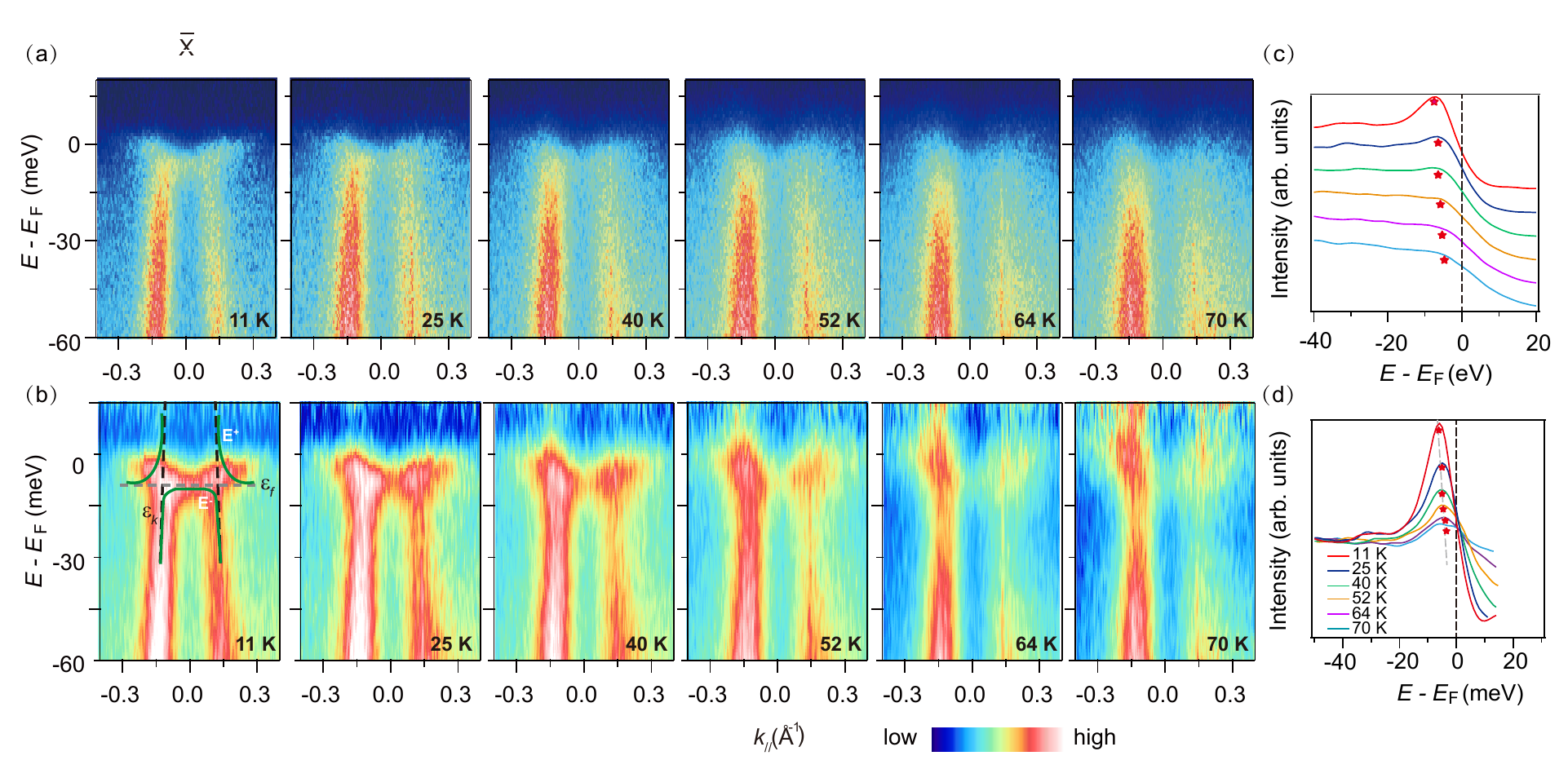}
\caption{Temperature dependence of the band structure around the $\bar{X}$  point. (a) Temperature dependence of the band structure around the $\bar{X}$ point. (b) Spectra in panel (a) after divided by the Fermi-Dirac function convolved with a Gaussian function representing the instrumental resolution to remove the thermal broadening contribution. (c) Temperature dependence of the EDCs around the $\bar{X}$  point in panel (a). (d) Temperature dependence of the EDCs around the $\bar{X}$ point in panel (b).}
\end{figure*}


We begin with a general characterization of different surface terminations of the cleaved URu$_2$Si$_2$ crystals. Figs. 2(a) and 2(b) show the valence band structure of different terminations of the freshly cleaved URu$_2$Si$_2$ crystals along the $\bar{\Gamma}$-$\bar{X}$ direction. Obvious differences can be found for these two terminations: i) One termination shows up an electron-like band around the $\bar{X}$ point, labelled by the star symbol in Fig. 2(a), while it is replaced by the hole-like band for another termination in Fig. 2(b). ii) High intensity of spectral weight can be found near $E_F$ labelled by a diamond symbol in Fig. 2(b), while it is weakened in Fig. 2(a). iii) A fast dispersive hole-like band with its top located at around 0.3 eV labelled by a triangle symbol can be found in Fig. 2(a), while it is almost absent in Fig. 2(b). These features can be used to distinguish different terminations in URu$_2$Si$_2$.

To further determine U- or Si- terminated surfaces, we performed band structure calculations of the two surfaces by DFT+DMFT method. The calculated band structures of U- and Si- terminations are displayed in Fig. 3(a) and 3(f), respectively. From theoretical calculations, a key signature of the two terminated surfaces is: an electron-like band around the $\bar{X}$ point in the U-terminated surface turns out to be a hole-like band for the Si-termination. This is consistent with our experimental results in Fig. 2. By comparing experimental data with theoretical calculations, it is clear that the band structure displayed in Fig. 2(a) is from the U-terminated surface, while Fig. 2(b) shows the band structure from the Si-terminated surface.

In order to show the detailed differences of U- and Si- terminations more clearly, APRES spectra of the two terminations near $E_F$ around the $\bar{\Gamma}$ and $\bar{X}$ points are displayed in Figs. 3(c-e) and Fig. 3(h-j), respectively. Meanwhile, photoemission intensity maps of the two terminations are displayed in Fig. 3(b) and 3(g). For the U-terminated surface, two hole-like features near the Fermi energy can be observed around the $\bar{\Gamma}$ point, labelled $\alpha$ and $\sigma$' in Fig. 3(c), respectively. These features can also be found from the calculation in Fig. 3(a). Among them, feature $\alpha$ is a hole-like band crossing $E_F$ at $\pm$0.2 ${\AA}^{-1}$, which contributes a hole pocket around the Brillouin zone center in Fig. 3(b). Feature $\sigma$' locates at around 30 meV below $E_F$, which is attributed to a surface state in previous ARPES results \cite{Andres.13}. These bands are also well reproduced in the calculations in Fig. 3(a). Around the $\bar{X}$ point, an electron-like band labelled $\eta$ can be found, which contributes to the electron-like pocket around the Brillouin zone corner and was well reproduced in the calculations in Fig. 3(a). For the Si-terminated surface, two hole-like features are observed around the $\bar{\Gamma}$ point, labeled $\omega$ and $\sigma$  in Fig. 3(h). Among them, feature $\omega$ crosses $E_F$ and contributes the hole-like pocket around the Brillouin zone center in Fig. 3(g). A hole-like band, named $\eta$' can be clearly observed around the $\bar{X}$ point, and contributes a hole-like pocket around the zone corner in Fig. 3(f). It is also noteworthy that the spot size of HeI lamp is rather large (around 1 mm), although in our experiments we have observed two sets of bands which display significantly different character, it is still possible that the data sets contain contributions from both cleaved planes and the $\sigma$' band observed for the U-terminated surface is a consequence of the strong intensity associated with the $\sigma$ band for the Si-terminated surface.

To further confirm the identification of different surface terminations, we perform STM measurements on the same samples immediately after ARPES measurements. The samples were transferred from ARPES chamber to STM chamber under ultra high vacuum condition very quickly. In Figs. 4(a, b), we show the typical topographic images of the samples that were used for ARPES measurements in Fig. 2(a). For these samples, two types of surfaces can be found, which we assigned as A and B respectively, see Figs. 4(a, b). The cleaved surface is dominated by surface A, and small portion of surface B can be found. The relative heights between different surfaces in Figs. 4(a) and (b) are displayed in Figs. 4(i) and 4(c), respectively.  By scanning the whole sample, we found that the probability of the occurrence of surfaces A and B is roughly 85\% and 15\%, respectively. The step size of the two surfaces is consistent with the relative height of U and Si layers, see Fig. 4(j). Fig. 4(d) displays the averaged electronic density of states of surface A, which is consistent with that of the U-terminated surface \cite{Aynajian10383}. Based on this, surfaces A and B can be assigned to U- and Si- terminated surfaces, respectively. These results are in line with ARPES measurements in Fig. 2(a) with the spectrum dominated by the U- terminated surface. Meanwhile, we found that the probability for Si- and U- terminated surfaces for the samples that were used for ARPES measurements in Fig. 2(b) is roughly 80\% and 20\%, see Figs. 4(e) and 4(f)), and the spectroscopy of the Si-terminated surface agrees well with previous STM results \cite{Aynajian10383}. Since the spot size of the He lamp is around 1 mm, the spectrum in Fig. 2(a) is dominated by the U-terminated surface but also with small contributions from the Si-terminated surface. This is the reason that there is still some residual intensity of the $\sigma$' band from ARPES spectrum in Fig. 3(c).

Having clearly characterized the different surface terminations in URu$_2$Si$_2$, we now turn our attention to the 5$f$-electron character in this compound. Previous ARPES results mainly concentrated on the relationship between the electronic structure and the ``hidden order" \cite{Andres.14,Andres.13,Andres.09,Kawasaki.11,Yoshida.12,Yoshida.10,Dakovski.11,Ito.99,Meng.13,PhysRevLett.110.186401}, and proposed that there is obvious change of the electronic structure during the ``hidden order" transition. Also, detailed temperature evolution of the electronic structure around the $\bar{\Gamma}$ point has been investigated, and a `M' shaped band was reported to be closely related to the ``hidden order" transition  \cite{Andres.14,Andres.13,Andres.09,Yoshida.12,Yoshida.10,Dakovski.11,Ito.99,PhysRevLett.110.186401}.
There is another key temperature scale in URu$_2$Si$_2$. According to  the resistivity data of URu$_2$Si$_2$, there is a rapid decrease below 55 K \cite{Maple.86,Palstra.85}, and it is proposed that the screening by light Ru-based $d$-electron bands of the $f$-electrons at each U atom apparently begins to alter the URu$_2$Si$_2$ electronic structure at this key temperature \cite{Schmidt.10}.  This behavior is similar to that of many hybridized $f$-electron rare-earth and actinide compounds. However, direct observation of the evolution of the electronic structure at this cross-over by ARPES is still lacking in URu$_2$Si$_2$. Here we extend the temperature range up to 70 K around the $\bar{X}$ point and focus on the evolution of the electronic structure at this key temperature around 55 K.


Figure 5 shows a detailed temperature evolution of the band structure around the $\bar{X}$ point from 11 K to 70 K of the Si-terminated surface. At 70 K, the photoemission is dominated by a strongly dispersive hole-like band and weak intensity around the $E_F$. Upon decreasing temperature, spectral weight near $E_F$ around the $\bar{X}$ point gradually increases and weakly dispersive hybridized bands can be observed around $E_F$. At 11 K, an obvious $f$-electron feature near $E_F$ can be clearly observed from the intensity plots, as demonstrated in Fig. 5(a), which indicates the hybridization between the $f$ band and conduction bands. The hybridization of this conduction band with the $f$ band causes the redistribution of the $f$ spectral weight and forms a weakly dispersive band near the $\bar{X}$ point. The $f$ spectral weight is significantly enhanced to the ``insight" of the two bands.  The hybridization of the $f$ band with the conduction band can be well described by a mean-field hybridization band picture, as illustrated by the dashed lines in Fig. 5(b), where $\varepsilon_f$ is the renormalized $f$-level energy, $\varepsilon_k$ is the conduction-band dispersion. The spectral weight of this hybridized band is gradually weakened as increasing temperature and becomes rather weak at 70 K. However, we did not observe any abrupt change of the electronic structure around the coherence temperature of 55 K \cite{Maple.86}. The $f$ spectral weight gradually weakens upon increasing temperature, still small part of $f$ spectral weight can be found above this temperature, which indicates that the $f$ electrons already start to hybridize with conduction electrons above the coherence temperature. This is in line with the quasiparticle scattering measurements \cite{Park.12}.
However, it is somewhat different from that found from the optical conductivity measurements, which showed that the hybridization almost starts at the coherence temperature around 55 K \cite{Nagel.12}. It is noteworthy that optical conductivity measurements are believed to be more bulk sensitive than ARPES, so the disagreement might originate from the surface states mainly detected by ARPES.

This temperature dependence of the electronic structure can be even more clearly observed from the spectra in Fig. 5(b) after divided by the resolution-convoluted Fermi-Dirac distribution at corresponding temperatures, and can also be reflected in the energy distribution curves (EDCs) at the $\bar{X}$ point in Figs. 5(c) and 5(d). From Figs. 5(c) and 5(d), the peak positions of the quasiparticles seem to move gradually towards $E_F$ upon increasing temperature, and this is similar with the evolution of the $f$-electron behavior in Ce- based compounds \cite{Qiuyun.17}. It is also noteworthy that we did not observe abrupt change of the electronic structure around the $\bar{X}$ point during the hidden-order transition. This is in line with previous ARPES results by Boariu $et~al.$ \cite{Andres.13}. They found the hidden-order parameter is anisotropic with pronounced changes at the $\bar{\Gamma}$ and $\bar{Z}$ points, while almost the same at the $\bar{X}$ point.
This gradually increased $f$ spectral weight with lowering temperature is similar with the 4$f$-electron behavior in the Ce-based compounds \cite{Qiuyun.17,CeIrIn5_qiuyun,CeRhIn5_qiuyun}.

\section{Conclusions}
In summary, we have discerned two well-defined and different types of spectra, which can be connected with the Si- and U-terminated surfaces of URu$_2$Si$_2$. In the U-terminated surface, an electron-like band is observed around the $\bar{X}$ point, which is replaced by the hole-like band for the Si-terminated surface. This can be a key signature to identify different surface terminations in URu$_2$Si$_2$. Meanwhile, obvious heavy quasiparticle bands can be observed at low temperature, and the strength of this band is gradually weakened as increasing temperature. We did not observe abrupt change of the electronic structure around the coherence temperature at the $\bar{X}$ point. Residual $f$ spectral weight can be found above the coherence temperature, which suggests the $f$ electrons start to hybridize with the conduction electrons above this temperature. Our results strongly suggest that the interaction between the lattice of heavy fermions and light conduction electrons plays a significant role during the whole process. Nonetheless, the relationship between the ``hidden order" and the interaction of the 5$f$ and conduction electrons remains an open question.

\begin{acknowledgments}
 We gratefully acknowledge enlightening discussions with P. Coleman and D. V. Vyalikh. This work is supported in part by the National Natural Science Foundation of China (Grants No. 11874330,11504342,11504341,11774320,11704347), Science Challenge Project (No.~TZ2016004), and National Key Research and Development Program of China (No.~2017YFA0303104). Some preliminary data were taken at the ARPES beam line of Shanghai Synchrotron Radiation Facility (SSRF, China).
\end{acknowledgments}


\end{document}